\documentclass[letterpaper,aps,pre,twocolumn,showcaps,superscriptaddress,amsmath,amssymb]{revtex4}
\usepackage{graphicx,color,natbib}

\newcommand{\pd}[2]{\frac{\partial #1}{\partial #2}}
\newcommand{\mean}[1]{\langle #1\rangle}

\newcommand{\sig}{\sigma}
\newcommand{\Om}{\Omega}

\newcommand{\Nc}{\mathcal N}
\newcommand{\Ns}{N_\text{LFS}}

\newcommand{\cv}{c_\text{v}}

\newcommand{\Tk}{T_\text{K}}
\newcommand{\Ts}{T_\text{s}}
\newcommand{\Tc}{T_\text{c}}
\newcommand{\tobs}{t_\text{obs}}

\begin{document}

\title{Liquid-liquid phase transition in an atomistic model glass former}

\author{Thomas Speck}
\affiliation{Institut f\"ur Physik, Johannes Gutenberg-Universit\"at Mainz,
  Staudingerweg 7-9, 55128 Mainz, Germany}
\author{C. Patrick Royall}
\affiliation{HH Wills Physics Laboratory, University of Bristol, Bristol BS8
  1TL, UK}
\affiliation{School of Chemistry, University of Bristol, Bristol BS8 1TS, UK}
\affiliation{Centre for Nanoscience and Quantum Information, Bristol BS8 1FD,
  UK}
\author{Stephen R. Williams}
\affiliation{Research School of Chemistry, Australian National University,
  Canberra, ACT 0200, Australia}

\begin{abstract}
  Whether the glass transition is related to an underlying thermodynamic
  singularity or whether it is a purely kinetic phenomenon is a major
  outstanding question in condensed matter physics. The main challenge one
  faces when studying supercooled liquids and glasses is of course that the
  dynamics of the constituent particles slows by many orders of
  magnitude. Thus it is hard to approach any possible transition with computer
  simulations, and one must extrapolate over many decades in relaxation time
  assuming no qualitative change in the system's behaviour. This limits the
  use of simulation, which would otherwise provide the detailed information
  with which one could discriminate between the numerous competing theoretical
  approaches. Here we make progress by introducing a sampling method that
  enables previously inaccessible dynamical regimes to be directly studied in
  an atomistic model glass former. Our study reveals a peak in the specific
  heat capacity, which we interpret as a drop in the density of states. Rather
  than being related to the formation of an ``ideal glass'', we provide
  evidence that the heat capacity peak is related to a liquid-liquid
  transition to a state rich in locally stable geometric motifs. This second
  liquid is structurally distinct from that observed at high temperature but
  remains amorphous.
\end{abstract}

%\pacs{64.70.Pf, 64.70.Ja}

\maketitle

% abbreviations used:
% LFS - locally favored structure
% LLT - liquid-liquid transition
% MD - molecular dynamics

%% ---- introduction ----

\section*{Introduction} 

The concept of a liquid-liquid transition (LLT) between
two distinct (metastable) supercooled liquids has attracted considerable
interest~\cite{tana99,mcmi07}. Some evidence for LLTs has been obtained in model
systems~\cite{elen10,ronc14} and metallic glass
formers~\cite{mcmi07,wei13}. Such transitions remain highly controversial both
in metallic systems~\cite{barn09} and molecular systems such as
water~\cite{limm11,palm14}. The application of thermodynamic concepts is,
  strictly speaking, not admissible due to the finite lifetime of both
  liquids, but might be a practical tool in situations where this lifetime
  vastly exceeds both typical measurement times and structural relaxation
  times.  Typical glass forming liquids often fall into this category.

At the glass transition the liquid falls out of equilibrium, in particular it
acquires a memory of its preparation. Although this event precludes a direct
observation, many theoretical approaches invoke an underlying thermodynamic
transition. Examples include the transition to an amorphous ground state (the
``ideal glass'')~\cite{cava09,bert11a} or a LLT to a state with the atoms
arranged in certain geometric motifs or \emph{locally favoured
  structures}~\cite{tana99,tarj05}. The pinning (or confinement) of
  particles has been argued to shift a thermodynamic transition into the
  accessible regime~\cite{camm10,bert13a}, and a loss of configurational
  entropy indicated by a peak in the heat capacity~\cite{mart14} has been
  reported.

Although convincing evidence for an underlying thermodynamic transition
remains elusive, it has been demonstrated recently that there is a
\emph{dynamical} transition in the space of \emph{trajectories}, \emph{i.e.},
sequences of configurations. Under biased sampling trajectories of sufficient
length undergo a first-order transition between an \emph{active} phase (the
supercooled liquid) and a dynamically \emph{inactive} phase, which has been
observed both in idealized lattice models~\cite{mero05} and in simulations of
atomistic model glass formers~\cite{hedg09,spec12b}. The latter case affords a
systematic computational means of preparing exceptionally stable glass
states~\cite{jack11,spec12b} (see also Ref.~\cite{sing13} for other methods to
prepare ultra-stable glasses). One suitable means to generate such transitions
in trajectory space is to employ biased sampling of trajectories based on an
order parameter characterizing \emph{time-averaged} populations of locally
favoured structures (LFS)~\cite{spec12b}. Connections between these inactive
phases prepared \textit{in silico} and those found in experiment (where there
is no bias) are only just beginning to be addressed~\cite{keys13}.

Configurations visited in the inactive phase have interesting
properties. While still amorphous, they are very stable with very large
structural relaxation times. Moreover, steepest descent quenches to
\emph{inherent states} indicate that the inactive phase probes configurations
very deep in the energy landscape relative to ``normal'' liquid
configurations~\cite{jack11,spec12b}. The latter observation opens the
possibility that these configurations have a finite weight also in the
equilibrium Boltzmann ensemble at a lower temperature. A major obstacle in
probing the link between deeply supercooled liquids and the inactive phase is
that in glass formers time scales are far beyond direct computational methods
such as molecular dynamics (MD) simulations. Here, to circumvent this problem
we employ a reweighting technique to gain insights into the behavior at very
low temperatures from data sampled at a mildly supercooled temperature. The
technique exploits the time-scale separation between vibrations and structural
relaxation~\cite{scio99}, whereby the relaxation time is nevertheless small
enough to allow efficient sampling. We find a peak in the specific heat
capacity $\cv$ at a temperature $T_\ast>\Tk$ higher than the extrapolated
Kauzmann temperature $\Tk$ corresponding to an ``ideal glass''
state. Moreover, we show that this peak is related to the transition in
trajectory space to the dynamical phase rich in LFS~\cite{spec12b}. From this
numerical evidence we conclude the existence of a LLT at $T_\ast$ from the
supercooled liquid to a state rich in LFS.

%% ---- method ----

\section*{Model} 

We consider the Kob-Andersen binary mixture~\cite{kob94}, a
well studied atomistic glass former that consists of 80\% large particles (A)
and 20\% small particles (B) interacting through truncated and shifted
Lennard-Jones pair potentials. We employ the original potential
parameters. All numerical values are reported in Lennard-Jones units with
respect to the large particles, and we set Boltzmann's constant to unity. We
study a system of $N=216$ particles at number density $N/V=1.2$ in a box with
constant volume $V$.

Our numerical scheme~\cite{spec12b} harvests trajectories with fixed number of
configurations $K$ at temperature $\Ts=0.6$ through a combination of
transition path sampling moves (shifting and half-shooting moves) to generate
trial trajectories~\cite{dell02}, the Metropolis criterion to accept or reject
trajectories, and replica exchange between quadratic biasing potentials of the
appropriate order parameter: the inherent state energy per particle $\phi$ and
the population $\Nc$ of particles in LFS, which couples to an external field
$\mu$. In this model, the LFS is a bicapped square
antiprism~\cite{cosl07,mali13a}.

In particular, we calculate the average of an observable $A$ at different
state points with
\begin{equation}
  \mean{A}(T,\mu) \equiv
  \frac{\mean{Ae^{-(1/T-1/\Ts)N\phi+\mu\Nc}}_0}{\mean{e^{-(1/T-1/Ts)N\phi+\mu\Nc}}_0}.
\end{equation}
The brackets $\mean{\cdot}_0$ denote the average over the sampled trajectories
using their unbiased equilibrium weight as indicated by the subscript. This
expression is evaluated employing the multistate Bennett acceptance ratio
method~\cite{shir08}. Assuming that inherent state energies and vibrational
free energy decouple (as appropriate here), it is straightforward to show that
averages of observables $A=A(\phi)$ that only depend on $\phi$ are equilibrium
averages in the absence of the field ($\mu=0$). Thus by reweighting according
to the inherent state energy, we access configurations corresponding to
temperatures beyond the range of normal simulations. See Methods for further
details.

%% ---- energy landscape ----

\begin{figure}[t]
  \centering
  \includegraphics{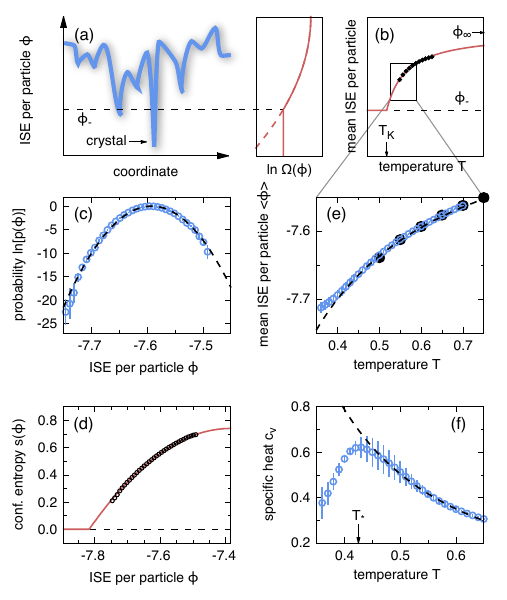}
  \caption{Statistical description of inherent states: (a)~Caricature of
    inherent state energies (ISE) $\phi$ along an arbitrary coordinate.
    Number of amorphous inherent states within a given energy interval is
    $\Om(\phi)$. Exponentially small $\Om(\phi)$ below $\phi_-$ (dashed
    continuation) corresponding to vanishing configurational entropy in the
    thermodynamic limit (solid line). (b)~Mean energy $\mean{\phi}$, which
    becomes equal to constant $\phi_-$ below $\Tk$. Symbols show inherent
    state energies from conventional MD simulations. (c)~Logarithm of the
    sampled probability distribution $p(\phi)$ for the sampling temperature
    $\Ts=0.6$. Dashed line is a Gaussian fit. (d)~Configurational entropy
    $s(\phi)$. The line is the expected quadratic behavior, the symbols are
    the data from (c) without temperature-bias. (e)~Mean energy $\mean{\phi}$
    as a function of temperature $T$ corresponding to the indicated range in
    (b). Shown are results from conventional MD simulations~($\bullet$) and
    results employing our reweighting scheme~($\circ$). Dashed line is the
    prediction Eq.~\eqref{eq:mean}. (f)~Heat capacity $\cv$ calculated from
    the inherent energy fluctuations exhibits a peak around
    $T_\ast\simeq0.425$. Dashed line is heat capacity for quadratic
    enumeration function.}
  \label{fig:landscape}
\end{figure}

\textbf{Energy landscape.} We first study the statistics of inherent
states~\cite{debe01}. Let $\Om(\phi)\propto\Om_\infty
e^{N\sig(\phi)}\delta\phi$ be the number of \emph{amorphous} inherent states
with energy per particle $\phi$ within an interval $\phi\pm\delta\phi/2$,
where $\sig(\phi)$ is the enumeration function. This scenario is sketched in
Fig.~\ref{fig:landscape}(a). Here, $\Om_\infty\simeq e^{Ns_\infty}$ is the
maximal available volume in configuration space. In the limit of large $N$,
the number $\Om$ is either extensive or becomes exponentially small. Hence, in
the thermodynamic limit, the extensive configurational entropy becomes
$\ln\Om(\phi)\simeq Ns(\phi)$ with $s(\phi)=s_\infty+\sig(\phi)$ for
$\phi\geqslant\phi_-$ and $s=0$ for extreme energies $\phi<\phi_-$. We assign
a temperature to the inherent structures based on $d\sig/d\phi=1/T$. First
pointed out by Kauzmann in 1948~\cite{kauz48}, the possibility that $\phi_-$
is reached at a finite temperature $\Tk$ has intrigued scientists ever since.
Figuratively speaking, at this temperature the system would ``run out'' of
amorphous configurations and would undergo a thermodynamic phase transition to
an ideal glas with constant inherent state energy $\phi_-$, see
Fig.~\ref{fig:landscape}(b).

In our simulations, we access the distribution $p(\phi)$ of inherent state
energies, which for $\Ts$ is plotted in Fig.~\ref{fig:landscape}(c). In
agreement with previous studies~\cite{buch99,scio99}, we find that this
distribution is well described by a Gaussian. Moreover, it has been
demonstrated that at low enough temperature (including the sampling
temperature $\Ts=0.6$ used here) the vibrational free energy is independent of
$\phi$ to a very good approximation~\cite{scio99}. We can thus extract the
quadratic enumeration function $\sig(\phi)=-(\phi-\phi_\infty)^2/J^2$ from the
measured distribution with fitted maximal energy $\phi_\infty\simeq-7.385$ and
scale $J\simeq0.502$. The resulting configurational entropy is plotted in
Fig.~\ref{fig:landscape}(d) using the value $s_\infty\simeq0.74$ reported
previously~\cite{scio99}. It demonstrates that our numerical scheme is able to
cover a wide range of inherent states and that the configurational entropy is
indeed very well described by a quadratic function. For the thermal average
one finds
\begin{equation}
  \label{eq:mean}
  \mean{\phi} = \phi_\infty - \frac{J^2}{2T} \qquad (\phi\geqslant\phi_-).
\end{equation}
Extrapolating the quadratic form for $\sig(\phi)$ to lower energies, one
easily derives $\Tk=J/(2\sqrt{s_\infty})$ for the Kauzmann temperature
yielding $\Tk\simeq0.29$ for the present system~\cite{scio99}. Whether such an
extrapolation is meaningful is debated, \emph{e.g.}, Stillinger has
pointed out that the ``melting'' of an ideal glass through defects would imply
$\Tk=0$~\cite{stil88}.

In Fig.~\ref{fig:landscape}(e) the mean inherent state energy $\mean{\phi}$ is
shown as a function of temperature. Down to $T=0.5$ we have run conventional
MD simulations and determined the average $\mean{\phi}$, which agrees well
with Eq.~\eqref{eq:mean}. Also plotted are the average inherent state energies
obtained through reweighting our simulation results obtained at $\Ts=0.6$ to
different temperatures, which agree with both the MD simulations and the
prediction Eq.~\eqref{eq:mean}. Strikingly, at lower temperatures we observe a
deviation from the behavior expected for a quadratic enumeration
function. This becomes even more pronounced when we consider the
susceptibility, \emph{i.e.}, the specific heat capacity
\begin{equation}
  \label{eq:cv}
  \cv = \pd{\mean{\phi}}{T} = \frac{\mean{\phi^2}-\mean{\phi}^2}{T^2}
\end{equation}
of the inherent states. From Eq.~\eqref{eq:mean} we expect
$\cv=\frac{1}{2}(J/T)^2$ for $T>\Tk$ and $\cv=0$ below the Kauzmann
temperature $\Tk$. As shown in Fig.~\ref{fig:landscape}(f), the heat capacity
drops below the expected behavior already for a temperature
$T_\ast\simeq0.425$ somewhat larger than the Kauzmann
temperature. 

%% ---- dynamical ----

\section*{Transitions in trajectory space} 

\begin{figure}[t]
  \centering
  \includegraphics{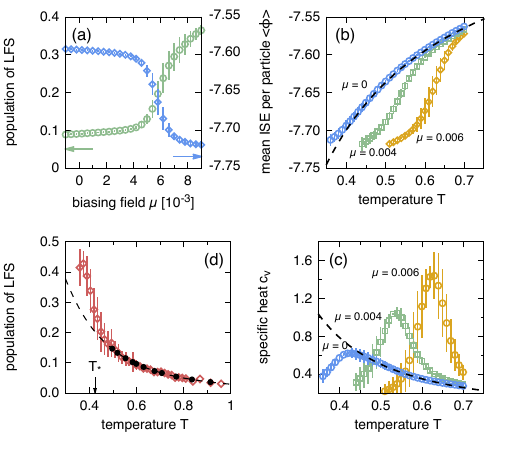}
  \caption{Transition in trajectory space: (a)~Mean population $n$ (left axis)
    and the mean inherent state energy $\mean{\phi}$ (right axis) as a
    function of $\mu$. Both observables show an abrupt change around
    $\mu_\ast\simeq5.8\times10^{-3}$. (b)~The mean energy $\mean{\phi}$ as a
    function of temperature extended to $\mu>0$,
    cf. Fig.~\ref{fig:landscape}(d). (c)~Specific heat for the same values of
    $\mu$, cf. Fig.~\ref{fig:landscape}(f). (d)~Population ($\bullet$) of LSF
    from conventional MD simulations. The dashed line is an extrapolation of
    this data using the phenomenological function
    $n=(1+(T/0.25)^{2.5})^{-1}$. In comparison, results ($\diamond$) from
    ``thermalized'' MD simulations show a jump, see text for further details.}
  \label{fig:traj}
\end{figure}

In order to elucidate the physical
origin of this drop in $\cv$, we now turn to the active-inactive
transition~\cite{spec12b} and make the conceptual leap from configurations to
trajectories. We spawn trajectories (symbolically denoted $x(t)$, where $t$ is
time) of length $\tobs=K\Delta t$ by integrating backward and forward in
time. To characterize trajectories, we count the number $\Nc[x(t)]$ of
particles that reside in LFS. In analogy to conventional thermodynamics, we
couple the order parameter $\Nc$ to an external field $\mu$ akin to a chemical
potential difference. We consider the perturbed distribution of
trajectories~\cite{hedg09}
\begin{equation}
  \label{eq:mu}
  P_\mu[x(t)] \propto P_0[x(t)]e^{\mu\Nc[x(t)]},
\end{equation}
which favors a large population of the structural motif for $\mu>0$. Although
we sample trajectories, for our analysis we only consider the initial and
central configurations, which we characterize in terms of the inherent state
energy $\phi$ as before and, in addition, the population $n=\mean{\Ns}/N$,
where $\Ns$ is the number of particles in LFS in the central configuration. In
Fig.~\ref{fig:traj}(a), both these quantities are plotted as a function of the
field $\mu$ at the sampling temperature $\Ts=0.6$. Around
$\mu_\ast\simeq5.6\times10^{-3}$, we observe a sudden change from the liquid
containing few LFS ($n\simeq0.09$) to a phase rich in LFS. At the same time,
the average inherent state energy drops from $\simeq-7.61$ to below $-7.7$,
which corresponds to a (unbiased) configurational temperature of less than
$0.4$, cf. Figs.~\ref{fig:landscape}(e) and~\ref{fig:traj}(b). The steepness
of the transition and the value of $\mu_\ast$ depend strongly on the
trajectory length $K$ and are compatible with a first-order
transition~\cite{spec12b}.

In order to relate the statistics of the potential energy landscape with the
dynamical transition, we extend our analysis by considering the trajectory
weight
\begin{equation}
  P_{\mu,T} \propto P_0 \exp\left\{ \mu\Nc-(1/T-1/\Ts)N\phi \right\}.
\end{equation}
This ``continuation'' allows us to study the system as a function of both
variables $\mu$ and $T$. In Fig.~\ref{fig:traj}(b) we show the average
inherent state energy $\mean{\phi}$ as a function of $T$ for three different
values of $\mu$. We observe a clear drop of the inherent state energy, which
is shifted to lower temperatures as $\mu$ is lowered. As expected from this
behavior, the heat capacity Eq.~\eqref{eq:cv} shown in Fig.~\ref{fig:traj}(c)
exhibits a peak that is also shifted to lower temperatures. As the temperature
is reduced, the peak height drops. This drop is consistent with a
dynamical critical point at $(\Tc>T_\ast,\mu>0)$ at which the
susceptibility may diverge~\cite{elma10}, see also Fig.~\ref{fig:phasedia}(b).

%% ---- linking ----

\section*{Linking the dynamical transition and $\cv$ peak} 

Extrapolating the
dynamical transition to $\mu=0$ suggests that the zero-bias $\cv$ peak would
also correspond to a transition to a (static) LFS-rich state. However,
studying this proposed thermodynamic LLT with MD simulations is not
straightforward due to the huge computational effort to properly equilibrate
the system at the required temperatures. Moreover, the scenario of a first
order transition implies that the liquid becomes metastable with respect to
the LFS-rich state (and possibly also to crystallisation). As a consequence,
the relevant configurations might not be found by straightforward molecular
dynamics. Still, these configurations contribute to equilibrium averages at
low temperatures due to their low energies, see Fig.~\ref{fig:traj}(d). Here
we have ``thermalized'' configurations harvested from the biased simulations
with conventional molecular dynamics at the configurational temperature $T$ of
the inherent state [inverting Eq.~\eqref{eq:mean}] for $t=150$, which is
sufficient for the system to undergo $\beta-$relaxation. Indeed, we see a
sharper rise in the LFS population upon cooling compared to the extrapolated
behaviour. Noting that since the LFS population is bounded to be $n<1$ it
appears that indeed there is a transition around $T_\ast\simeq0.425$ to an
LFS-rich state.

\begin{figure}[t]
  \centering
  \includegraphics{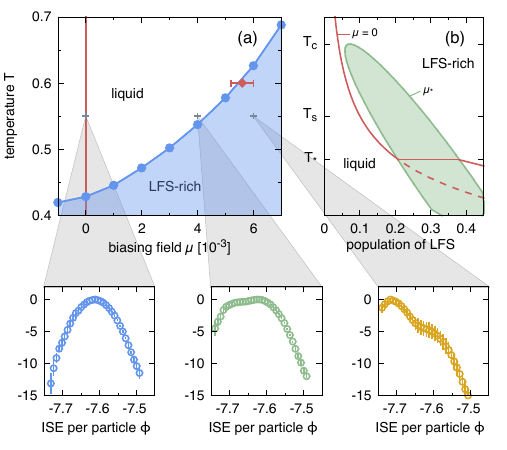}
  \caption{Phase diagram: (a)~The symbols ($\bullet$) show where the specific
    heat capacity $\cv$ reaches its maximum as a function of both $T$ and
    $\mu$ for $K=100$ (the line is a guide to the eye). The single diamond
    indicates the position of the dynamic phase transition and its statistical
    uncertainty. Shown below for $T=0.55$ are probability distributions
    $\ln[p(\phi)]$ of the inherent state energy per particle for the three
    indicated values of the biasing field (dynamical chemical potential)
    $\mu$. One distinguishes two populations at different inherent state
    energies, which we ascribe to the liquid and LFS-rich phases,
    respectively. (b)~Sketch of the corresponding $n$-$T$ plane: Below the
    critical temperature $\Tc$ phase coexistence is possible (shaded area
    bounded by $\mu_\ast$). The solid red line corresponds to $\mu=0$, where
    the dashed line indicates the metastable branch below $T_\ast$. At
    $T_\ast$ one has $\mu_\ast=0$ and a thermodynamic LLT occurs.}
  \label{fig:phasedia}
\end{figure}

In Fig.~\ref{fig:phasedia}(a) we present the phase diagram between the
supercooled liquid and the LFS-rich phase as signaled by the maximum of the
specific heat. Also shown are distributions of the inherent state energies at
configurational temperature $T=0.55$ for three values of $\mu$. Close to
coexistence one discerns two populations having different inherent state
energies that reverse their relative weights passing through the phase
boundary. This observation is again compatible with the scenario of a phase
transition. Note that for $\mu\neq0$ the position of the phase boundary and
also the peak value of the heat capacity depends on the value of $K$ used to
sample trajectories. However, for two trajectory lengths $K=100$ and $K=60$ we
have checked that for $\mu=0$ both $T_\ast$ and the value of $\cv$ agree as
expected. Finally, in Fig.~\ref{fig:phasedia}(b) a sketch of the corresponding
$n$-$T$ plane is shown. Below a critical temperature, dynamical coexistence of
liquid and LFS-rich phase is possible. Unconstrained equilibrium dynamics
corresponds to a vanishing field $\mu=0$, which lies in the liquid region as
long as $\mu_\ast>0$. The observed jump in the population suggests that at
$T_\ast$ a LLT occurs, whereby $\mu_\ast=0$.

%% ---- conclusions ----

\section*{Conclusions} 

We have developed a reweighting technique that allows us
to access a wide range of inherent states and thus exceptionally low
temperatures in a model glass forming liquid. This opens a perspective of what
actually happens as the system starts to run out of configurational entropy.
We find a peak in the specific heat, which occurs at a temperature
$T_\ast\simeq0.425$ above the Kauzmann temperature $\Tk\simeq0.29$ (estimated
under the condition that the quadratic behavior of the enumeration function
continues to lower energies~\cite{scio99}). This transition is distinct from
the dynamical crossover to an energy landscape dominated regime since we
sample configurations and determine their weight solely based on their
energy. In other words, our reweighting is not affected by sampling
limitations that usually prevent simulation techniques from sampling lower
temperatures.

The connection with a dynamical phase transition at a higher temperature to a
state rich in locally favoured structures enables insights into the nature of
the transition underlying the $\cv$ peak. The dynamical transition, which
occurs under a non-zero field controlling the chemical potential difference
between liquid and LFS particles, also features a $\cv$ peak, and upon
reducing the chemical potential we find a line of maxima connecting both
transitions. We thus interpret this line as the phase boundary corresponding
to a liquid-liquid transition between a low-temperature LFS-rich phase and the
normal supercooled liquid. Such a transition is in accord with the geometric
frustration approach~\cite{tarj05}.

While an LLT will have significant consequences for the glass forming
properties of the studied model, we have not attempted to identify a glass
transition line, which will depend strongly on time scales. An important
question lies in the generality of our findings. We have considered perhaps
the most popular atomistic glassformer. Our findings should certainly be
checked with other models.  The Kob-Andersen model we have considered is based
on a metallic glass, NiP. Indeed a number of metallic glassformers are known
to undergo LLTs~\cite{mcmi07,wei13}.  Regarding the glassforming properties we
note links between LLTs and fragile-to-strong transitions in both metallic
~\cite{maur10} and molecular~\cite{mall10} glassformers. If such a connection
between LLTs and strong low-temperature liquids can be established, the
consequences for a thermodynamic glass transition are profound as in the limit
of a perfectly strong liquid there is no transition.

%% ---- methods ----

\section*{Methods} Trajectories are stored as sequences
$X=(x_{-K/2},\dots,x_{K/2})$ of $K+1$ configurations $x_i$. We use a
stochastic thermostat to control temperature, \emph{i.e.}, we draw all
particle velocities from a Maxwell-Boltzmann distribution before integrating
for a time $\Delta t=1.5$ using the velocity Verlet algorithm and storing the
next configuration. In addition, inherent states $\hat x_i$ are obtained by a
minimization of the potential energy $\Phi(x)$ of configurations using the
FIRE algorithm~\cite{bitz06}. To make the harvesting of millions of
trajectories feasible, we limit the number of FIRE iterations to 1,000. Hence,
the reported values for $\phi=\Phi/N$ should be understood as upper (although
tight) bounds to the true inherent state energies.

We perform two independent sets of biased simulations using quadratic bias
functions of the relevant observable that are spaced equidistantly. First, we
employ the total number $\Nc[X]\equiv\sum_{i=-K/2}^{K/2}\sum_{k=1}^N h_k(x_i)$
of particles participating in LFS motifs in the whole trajectory, where
$h_k(x)$ is an indicator function that is unity if particle $k$ of
configuration $x$ is part of an LFS and zero otherwise~\cite{spec12b}. Second,
we bias trajectories using the inherent state energy $\phi$. For each umbrella
we harvest 120,000 trajectories in the first run and 60,000 trajectories in
the second run and combine the data. To estimate errors, the data is split
into three sets and analyzed independently.

%% ---- acknowledgments ----

\textbf{Acknowledgments.} CPR acknowledges the Royal Society and European
Research Council (ERC consolidator grant NANOPRS, project number 617266). This
work was carried out using the computational facilities of the Advanced
Computing Research Centre, University of Bristol. We thank L. Berthier,
D. Chandler, P. Harrowell, J. Garrahan, R. Jack  and H. Tanaka for discussions and
helpful comments.

%% ---- bibliography ----

%\bibliographystyle{naturemag}
%\bibliography{refs}

\end{document}